\documentclass[a4paper, 12pt, conference]{ieeeconf}  % Comment this line out
\IEEEoverridecommandlockouts                              % This command is only
\usepackage{datetime}
\usepackage{graphicx} % for pdf, bitmapped graphics files
\usepackage{mathptmx} % assumes new font selection scheme installed
\usepackage{times} % assumes new font selection scheme installed
\usepackage{amsmath} % assumes amsmath package installed
\usepackage{amssymb}  % assumes amsmath package installed
\newtheorem{lem}{Lemma}
\newtheorem{thm}{Theorem}

\def\<{\leqslant}           % nice less than or equal to sign
\def\>{\geqslant}           % nice larger than or equal to sign
\def\d{\partial}

\def\Re{{\rm Re}}   % real part
\def\Im{{\rm Im}}   % imaginary part

\def\cH{{\cal H}}   % Hardy space
\def\mA{{\mathbb A}}    % space of real antisymmetric matrices
\def\mR{{\mathbb R}}    % real line
\def\Tr{{\rm Tr}}       % matrix trace
\def\rT{{\rm T}}        % matrix transpose
\def\bE{{\bf E}}    % expectation

\def\[[[{[\![\![}
\def\]]]{]\!]\!]}

\def\bra{{\langle}}
\def\ket{{\rangle}}

\def\BRA{\Big\langle}
\def\KET{\Big\rangle}

\def\re{{\rm e}}        % number e
\def\rd{{\rm d}}        % differential

\def\cL{{\mathcal L}}

\def\br{{\bf r}}
\def\x{\times}
\def\ox{\otimes}

\def\cZ{{\mathcal Z}}

\def\cF{{\cal F}}
\def\cW{{\mathcal W}}

\def\cX{{\mathcal X}}

\def\cC{{\cal C}}

\def\cA{{\cal A}}
\def\cB{{\cal B}}
\def\cE{{\mathcal E}}

\def\mS{{\mathbb S}}

\def\Ups{\Upsilon}

\title{\Large \bf
Coherent Quantum Filtering for Physically Realizable Linear Quantum Plants$^*$
}

\author{Igor G. Vladimirov, \qquad Ian R. Petersen% | \today, \currenttime
\thanks{$^*$This work is supported by the
Australian Research Council. The authors are with the School of Engineering and Information Technology, University of New South Wales at the Australian Defence Force Academy, Canberra, ACT 2600, Australia: {\tt igor.g.vladimirov@gmail.com, i.r.petersen@gmail.com}.
}
}
\onecolumn
\begin{document}

\maketitle

\thispagestyle{empty}
%\pagestyle{empty}

%%%%%%%%%%%%%%%%%%%%%%%%%%%%%%%%%%%%%%%%%%%%%%%%%%%%%%%%%%%%%%%%%%%%%%%%%%%%%%%%%%%%%%%%%%%%%%%%%%%
\begin{abstract}
The paper is concerned with a problem of coherent (measurement-free) filtering for physically realizable (PR) linear quantum plants. The state variables of such systems satisfy canonical commutation relations %(CCRs)
and are governed by linear quantum stochastic differential equations, dynamically equivalent to those of an open quantum harmonic oscillator. The problem is to design another PR quantum system,  connected unilaterally  to the output of the plant and playing the role of a quantum filter, so as to minimize a mean square discrepancy between the dynamic variables of the plant and the output of the filter. This coherent quantum filtering (CQF)
formulation is a simplified feedback-free version of the coherent quantum LQG control problem which remains open
despite recent  studies.
%We establish a lower bound for the mean square estimation error by using an algebraic Riccati equation with complex coefficients which takes into account both the dynamics of the quantum plant and
%the CCRs whose preservation is part of the PR conditions.
The CQF problem is transformed into a constrained covariance control problem which is  treated by using  the Frechet differentiation of an appropriate Lagrange function with respect to the matrices of the filter.
%A salient feature of the optimal filter in comparison with the classical Kalman filter is the dependence of the solution on the weight matrix which specifies the quadratic performance index.
%We also discuss a ``counter-commutation'' technique for the CQF design which allows the Heisenberg uncertainty principle to be eliminated from affecting the signal being minimized. Finally, Peres-Horodecki-Simon separability criterion is applied to investigate the quantum entanglement between the state variables of the plant and the optimal quantum filter.

%The associated Hamilton-Jacobi-Bellman equation  for the minimum cost function involves  Frechet differentiation with respect to matrix-valued variables.
%The gain matrices of the CQLQG optimal  controller  are shown to satisfy a quasi-separation property as a weaker quantum counterpart of the filtering/control decomposition of classical LQG controllers.
%Although such controllers are also inherently noisy,  the use of coherent control avoids the loss of quantum information caused by classical measurements.  The preservation of canonical commutation relations for the state of the closed-loop system is shown to cover the separate PR conditions for the plant and controller.
%The algorithm is demonstrated for a quantum-optical example of stabilizing an atom trapped in a cavity.

\end{abstract}

%%%%%%%%%%%%%%%%%%%%%%%%%%%%%%%%%%%%%%%%%%%%%%%%%%%%%%%%%%%%%%%%%%%%%%%%%%%%%%%%%%%%%%%%%%%%%%%%%%%
\section{Introduction}
%%%%%%%%%%%%%%%%%%%%%%%%%%%%%%%%%%%%%%%%%%%%%%%%%%%%%%%%%%%%%%%%%%%%%%%%%%%%%%%%%%%%%%%%%%%%%%%%%%%

Interconnection of open systems, whose internal dynamics are affected by interaction with the surroundings, is often engineered so as to stabilize the resulting network of such systems via redistribution and dissipation of energy generated  by active nodes or coming from the environment. In addition to its role in redirecting the energy flow, interaction provides a universal mechanism for creating correlations whereby current states of different subsystems acquire and store dynamic ``footprints'' of each other  and of the past history of the whole system.
This informational aspect of interaction is directly employed in the classical Kalman filter whose state is continuously updated by the measurement process from a stochastic system,  thus enabling  the filter to develop and maintain relatively strong correlation with the unknown state of the system. The ability of such a filter to track a classical linear system
with finitely many degrees of freedom
and known dynamics by extracting as much information from the noisy observations as possible is, in principle,  limited only by digital implementation.
The situation is qualitatively different in regard to estimating the dynamic variables of a quantum stochastic system which are noncommutative operators  on a Hilbert space evolving in time according to the laws of quantum mechanics \cite{M_1998}. The measurements, which result from the interaction of the quantum system with a  relatively invasive classical device, % (even if they are subsequently processed on an ideal computer),
are accompanied by irreversible loss of quantum information as a consequence of the projection postulate of quantum mechanics \cite{H_2001}. It is the idea of using a system of the same kind, that is, another quantum system, weakly coupled to the quantum mechanical object of interest (a quantum plant), which underlies the coherent (that is, measurement-free) quantum filtering/control paradigm. This approach replaces  measurement with interaction  of quantum systems, possibly mediated by light fields, where the energy flow can be employed for stabilization/control \cite{JG_2010} and the quantum information manifests itself through quantum correlations %being developed
between the dynamic variables of the systems in the course of time.
An important class of quantum stochastic systems is provided by open quantum harmonic oscillators \cite{EB_2005,GZ_2004} whose variables satisfy canonical commutation relations (CCRs) and are governed by linear quantum stochastic differential equations (QSDEs) driven by boson fields \cite{P_1992}. In combination with the preservation of CCRs under unitary evolutions in the Heisenberg picture of quantum dynamics,
 the specific energetics of such systems (quadratic Hamiltonian and linear coupling to the external fields) imposes certain constraints \cite{JNP_2008,SP_2012}  on the coefficients of the governing linear QSDEs in order for them to be physically realizable (PR) as open quantum harmonic oscillators. Such systems can be implemented in practice  by using quantum-optical components \cite{GZ_2004}. Being quadratic with respect to the state-space matrices,  the PR constraints make the coherent quantum counterpart \cite{NJP_2009}  of the classical LQG control problem \cite{KS_1972} substantially harder to solve than its classical predecessor. In fact, the coherent quantum LQG (CQLQG) feedback design problem remains open despite recent studies \cite{VP_2011a,ZJ_2011} which explore different approaches to its solution.
In the present paper, we consider an infinite-horizon coherent quantum filtering (CQF) problem for PR linear quantum plants. The question of interest is to design another PR quantum system,  connected unilaterally  to the output of the plant and playing the role of a quantum filter, so as to minimize a steady-state mean square discrepancy between the dynamic variables of the plant and the output of the filter.
%We establish a lower bound for the mean square estimation error by using an algebraic Riccati equation with complex coefficients which takes into account both the dynamics of the quantum plant and
%the CCRs whose preservation is part of the PR conditions.
In the absence of measurements and in the presence of PR constraints,  the machinery of recursive Bayesian estimation (including conditional expectations), so useful for Kalman filtering, is inapplicable to the CQF problem.
Following \cite{VP_2011a} based on algebraic ideas from \cite{BH_1998,SIG_1998}, we transform the CQF problem into a constrained covariance control problem which is  treated by using  the Frechet differentiation of an appropriate Lagrange function with respect to the matrices of the filter. 
Since the CQF setting is a simplified feedback-free version of the CQLQG control problem mentioned above, this leads to a more explicit set of algebraic equations for the state-space matrices of an optimal PR quantum filter which are amenable to further analysis to be published elsewhere. Note that the recognition of the need to take into account  PR constraints  in coherent quantum filtering problems dates back to \cite{GS_1970}, although that work was mainly concerned with measurement-based mean square optimal filtering. A coherent quantum filtering problem  has recently been discussed in \cite{MJ_2012}, where, unlike the present paper,  optimization of the filter was not considered.
\section{Physically Realizable Quantum Plant}\label{sec:plant}
%%%%%%%%%%%%%%%%%%%%%%%%%%%%%%%%%%%%%%%%%%%%%%%%%%%%%%%%%%%%%%%%%%%%%%%%%%%%%%%%

As in the linear quantum control settings \cite{JNP_2008,NJP_2009,VP_2011a} mentioned above,   the quantum plant considered below is an open quantum stochastic system with canonically commuting variables whose internal dynamics are affected by the quantum noise from the environment.
More precisely, the  plant has $n$ dynamic variables $x_1(t), \ldots, x_n(t)$ which are self-adjoint operators on the tensor product Hilbert space\footnote{Some of the spaces and parameters  associated with the plant are equipped with the subscript ``$1$'', whereas the subscript ``$2$'' is used for analogous objects pertaining to the quantum filter introduced in Section~\ref{sec:filter}.} $\cH_1\ox\cF_1$ evolving in time $t$ and satisfying the canonical commutation relations (CCRs)
\begin{equation}
\label{xCCR}
    [x,x^{\rT}]
    :=
    ([x_j, x_k])_{1\< j,k\< n}
    =
    xx^{\rT} - (xx^{\rT})^{\rT}
    =
    2i \Theta_1.
\end{equation}
Here, $x:= (x_k)_{1\< k \< n}$ is the vector\footnote{Vectors are organized as columns unless specified otherwise, and the transpose $(\cdot)^{\rT}$ acts
on matrices with operator-valued entries as if the latter were scalars.} of the plant variables (the time argument is often omitted for the sake of brevity), $[\eta,\zeta]:= \eta\zeta - \zeta\eta$ is the commutator of operators, $i:= \sqrt{-1}$ is the imaginary unit, and $\Theta_1$ is a constant real antisymmetric matrix of order $n$ (the subspace of such matrices is denoted by $\mA_n$) which is assumed to  be nonsingular, so that $n$ is even.  The operators $x_1(0), \ldots, x_n(0)$ act on the initial complex separable Hilbert space $\cH_1$ of the system, and $\cF_1$ is the boson Fock space \cite{P_1992}, which provides a domain for the action of the quantum Wiener processes $w_1(t), \ldots, w_{m_1}(t)$. The latter are self-adjoint operators on $\cF_1$ (which are obtained from pairs of field annihilation and creation operators by using complex unitary  $(2\x 2)$-matrices) and   represent the quantum noise from the environment with the quantum Ito table
%\begin{equation}
%\label{wtable}
$        \rd w \rd w^{\rT}
    =
    \Omega_1\rd t
$. %\end{equation}
Here,
$w:= (w_k)_{1\< k\< m_1}$, and $\Omega_1$ is a constant complex positive semi-definite  Hermitian matrix of order $m_1$. Similarly to (\ref{xCCR}),  the imaginary part
$J_1 := \Im \Omega_1 \in \mA_{m_1}$  of the quantum Ito matrix $\Omega_1$
  specifies the CCRs between the quantum Wiener processes as
\begin{equation}
\label{wCCR}
        [
            \rd w, \rd w^{\rT}
        ]
    =
    2i J_1\rd t.
\end{equation}
In what follows, the real part $\Re \Omega_1$ is the identity matrix $I_{m_1}$ of order $m_1$, so that
\begin{equation}
\label{Om1}
    \Omega_1
    =
    I_{m_1} + iJ_1.
\end{equation}
Also, it is assumed that the noise dimension $m_1$ is even, and  the CCR matrix $J_1$ has a canonical form
\begin{equation}
\label{canJ1}
    J_1
    :=
    {\small\begin{bmatrix}
    0 & 1\\
    -1 & 0
    \end{bmatrix}}
    \ox
    I_{\mu_1}
    =
    {\small\begin{bmatrix}
    0 & I_{\mu_1}\\
    -I_{\mu_1} & 0
    \end{bmatrix}},
\end{equation}
where $\ox$ is the Kronecker product of matrices, and $\mu_1:= m_1/2$.
%
%
%Also, $(\cdot)^{\dagger}:= ((\cdot)^{\#})^{\rT}$ denotes the transpose
%of the entry-wise adjoint $(\cdot)^{\#}$. In application to complex matrices,  $(\cdot)^{\dagger}$ reduces to the complex conjugate transpose $(\overline{(\cdot)})^{\rT}$ and is written as $(\cdot)^*$.
%
%Here, , $I_m$ is the identity matrix of order $m$ (the subscript will often be omitted for the sake of brevity), and $J_1$ is a real antisymmetric matrix, which is given by
%and specifies the  canonical commutation relations (CCRs) for the quantum noise of the plant as
%
%
%
%a complex separable Hilbert space $\mH_1$  assembled into a vector.
The plant state vector $x$ evolves  in time and contributes to a $p$-dimensional output of the plant  $y$ (whose entries are also self-adjoint operators on $\cH_1\ox \cF_1$) according to QSDEs
\begin{equation}
\label{x_y}
    \rd x
    =
    A x\rd t  +  B \rd w,
    \qquad
    \rd y
    =
    C x\rd t  +  D \rd w.
\end{equation}
Here,
$
    A\in \mR^{n\x n}
$,
$
    B\in \mR^{n\x m_1}
$,
$
    C\in \mR^{p\x n}
$,
$
    D\in \mR^{p\x m_1}
$ are given constant matrices, with $A$ Hurwitz.
In addition to the asymptotic stability,
the plant is assumed to be physically realizable (PR) as an open quantum harmonic oscillator \cite{EB_2005,GZ_2004} whose Hamiltonian is quadratic and the coupling operators are linear with respect to the plant variables. By the results of \cite{JNP_2008,NJP_2009,SP_2012}, in the case of linear  quantum dynamics being considered, the PR property is equivalent to the algebraic relations
\begin{align}
\label{plantPR1}
    A\Theta_1 + \Theta_1 A^{\rT} + BJ_1 B^{\rT} &= 0,\\
\label{plantPR2}
    \Theta_1C^{\rT} + BJ_1D^{\rT} &=0
\end{align}
which describe the preservation in time of the CCRs between the state and output variables of the plant described by (\ref{xCCR}) and $[x,y^{\rT}] = 0$. Indeed, (\ref{plantPR1}), (\ref{plantPR2}) are obtained from the relationships
\begin{align}
\label{xx}
    \rd [x,x^{\rT}] & = (A[x,x^{\rT}] + [x,x^{\rT}]A^{\rT} +2iBJ_1 B^{\rT})\rd t,\\
\label{xy}
    \rd [x,y^{\rT}] & = (A [x,y^{\rT}] + [x,x^{\rT}]C^{\rT} + 2iBJ_1 D^{\rT}) \rd t
\end{align}
which follow from the bilinearity of the commutator \cite{M_1998}, the quantum Ito rule, and the commutativity between adapted processes  and forward increments of the quantum Wiener process \cite{P_1992},  in view of (\ref{wCCR}), (\ref{x_y}).

%%%%%%%%%%%%%%%%%%%%%%%%%%%%%%%%%%%%%%%%%%%%%%%%%%%%%%%%%%%%%%%%%%%%%%%%%%%%%%%%
\section{Physically Realizable Coherent Quantum Filter}\label{sec:filter}
%%%%%%%%%%%%%%%%%%%%%%%%%%%%%%%%%%%%%%%%%%%%%%%%%%%%%%%%%%%%%%%%%%%%%%%%%%%%%%%%

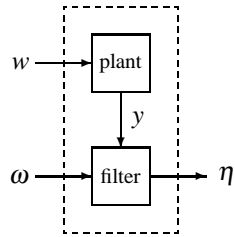
\begin{figure}[htpb]
\centering
\unitlength=0.75mm
%\linethickness{0.6pt}
\begin{picture}(30.00,43.00)
    \put(5,5){\dashbox(20,40)[cc]{}}
    \put(10,30){\framebox(10,10)[cc]{{\scriptsize plant}}}
    \put(10,10){\framebox(10,10)[cc]{{\scriptsize filter}}}
    \put(0,15){\vector(1,0){10}}
    \put(0,35){\vector(1,0){10}}
    \put(20,15){\vector(1,0){10}}
    \put(-1,35){\makebox(0,0)[rc]{{\small$w$}}}
    \put(-1,15){\makebox(0,0)[rc]{{\small$\omega$}}}
    \put(32,15){\makebox(0,0)[lc]{{\small$\eta$}}}
    \put(17,25){\makebox(0,0)[lc]{{\small$y$}}}
   \put(15,30){\vector(0,-1){10}}
\end{picture}\vskip-5mm
\caption{The dynamics of the filter are influenced by the quantum Wiener process $\omega$ and the plant output $y$ through a unilateral cascade connection  described by (\ref{x_y}), (\ref{xi_eta}).
}
\label{fig:cascade}
%\end{center}
\end{figure}

Consider another PR open quantum stochastic system with $q$ canonically commuting dynamic variables which are influenced by the environment and the plant in a unilateral fashion. Due to this cascade connection,  which is shown in Fig.~\ref{fig:cascade},
%==============================================================================
%==============================================================================
the state and output variables of the second system under consideration acquire quantum correlation with the plant variables in the course of time, which enables this system to be regarded as a coherent (that is, measurement-free) quantum filter. A performance criterion for such a filter  will be specified in Section \ref{sec:problem}. Now, the filter state variables $\xi_1(t), \ldots, \xi_q(t)$ are assumed to be self-adjoint operators on the Hilbert space $\cH_1\ox \cH_2 \ox \cF_1\ox \cF_2$ satisfying  the CCRs
\begin{equation}
\label{xiCCR}
    [\xi,\xi^{\rT}]
    =
    2i \Theta_2,
\end{equation}
where $\xi:= (\xi_k)_{1\< k \< q}$ and $\Theta_2\in \mA_q$, with $q$ even and $\det \Theta_2 \ne 0$. Here, $\cH_2$ is the initial space of the filter and $\cF_2$ is the boson Fock space which provides a domain for the action of a quantum Wiener process $\omega:= (\omega_k)_{1\< k\< m_2}$. The latter commutes with the plant noise $w$ and drives the filter variables according to the QSDE
\begin{equation}
\label{xi_eta}
    \rd \xi
     =
    a\xi\rd t + b \rd \omega + e\rd y,
    \qquad
    \rd \eta
     =
    c\xi\rd t + d \rd \omega,
\end{equation}
where
$
    a \in \mR^{q\x q}
$,
$
    b\in \mR^{q\x m_2}
$,
$
    c \in \mR^{r\x q}
$,
$
    d\in \mR^{r\x m_2}
$,
$
    e\in \mR^{q\x p}
$
are constant matrices, and $a$ is Hurwitz.  The quantum Wiener process $\omega$ is assumed to have even dimension $m_2$ and a canonical quantum Ito table
\begin{equation}
\label{Om2}
    \rd \omega \rd \omega^{\rT}
    =
    \Omega_2 \rd t,
    \qquad
    \Omega_2:=
    I_{m_2}+iJ_2,
\end{equation}
analogous to (\ref{Om1}), (\ref{canJ1}). Here,
$J_2\in \mA_{m_2}$ is a CCR matrix of the filter noise in the sense that
\begin{equation}
\label{J2}
    [\rd \omega, \rd \omega^{\rT}]
    =
    2iJ_2\rd t,
    \qquad
    J_2
    :=
    {\small\begin{bmatrix}
    0 & I_{\mu_2}\\
    -I_{\mu_2} & 0
    \end{bmatrix}},
\end{equation}
with $\mu_2:= m_2/2$.
The matrices $b$, $e$ in (\ref{xi_eta}) play the role of gain matrices of the quantum filter with respect to the filter noise $\omega$ and the plant output $y$. The plant, filter and the environment form a closed quantum system which is governed by
 (\ref{x_y}), (\ref{xi_eta}). Therefore, the $(n+q)$-dimensional vector
\begin{equation}
\label{cX}
\cX :=
    {\small\begin{bmatrix}
        x\\
        \xi
    \end{bmatrix}},
\end{equation}
formed by the plant and filter state variables, is driven by the combined quantum Wiener process
\begin{equation}
\label{cW}
\cW :=
    {\small\begin{bmatrix}
        w\\
        \omega
    \end{bmatrix}}
\end{equation}
of dimension $m:=m_1 + m_2$
according to the QSDE
\begin{equation}
\label{closed}
    \rd \cX
      =
      \cA     \cX \rd t +   \cB      \rd \cW,
\end{equation}
with
\begin{equation}
\label{cAB}
    \cA
    :=
        {\small\begin{bmatrix}
                  A       & 0\\
                  e C   & a
        \end{bmatrix}},
    \qquad
    \cB
    :=
        {\small\begin{bmatrix}
                   B       & 0\\
                  e D   & b
        \end{bmatrix}}.
\end{equation}
The matrices $\cA$, $\cB$ have a block lower triangular structure due to the absence of feedback,  as opposed to the closed-loop quantum control settings \cite{JNP_2008,NJP_2009,VP_2011a}. The plant and filter noises $w$, $\omega$ result from interaction of the systems with external boson fields which are assumed to be in the product vacuum state
%\begin{equation}
%\label{ups}
$    \upsilon
    :=
    \upsilon_1\ox \upsilon_2
$ %\end{equation}
on the  space $\cF_1\ox \cF_2$. Since  the noises commute with each other and are uncorrelated, then, in view of (\ref{wCCR})--(\ref{canJ1}) and (\ref{Om2}), (\ref{J2}), the combined Wiener process $\cW$ in (\ref{cW}) has a block diagonal quantum Ito table
\begin{equation}
\label{cWtable}
        \rd \cW  \rd \cW ^{\rT}
    =
    \Omega\rd t,
    \qquad
    \Omega
    :=
    {\small\begin{bmatrix}
        \Omega_1 & 0\\
        0 & \Omega_2
    \end{bmatrix}}
    =
    I_m+iJ.
\end{equation}
Here, $J\in \mA_m$ denotes the corresponding CCR matrix of $\cW$:
\begin{equation}
\label{J}
    [\rd \cW, \rd \cW^{\rT}]
    =
    2iJ\rd t,
    \qquad
    J
    :=
    {\small\begin{bmatrix}
    J_1 & 0\\
    0 & J_2
    \end{bmatrix}}.
\end{equation}
In a similar vein, the plant and filter variables are assumed to commute, so that the combined vector  (\ref{cX})  has a block-diagonal CCR matrix:
\begin{equation}
\label{cXCCR}
    [\cX, \cX^{\rT}]
    =
    2i
    \Theta,
    \qquad
    \Theta:=
    {\small\begin{bmatrix}
        \Theta_1 & 0\\
        0 & \Theta_2
    \end{bmatrix}},
\end{equation}
where use is made of (\ref{xCCR}), (\ref{xiCCR}). Due to the unitary evolution of the isolated  system formed by the plant, filter and their environment, the CCR matrix $\Theta$ is also preserved in time, which is equivalent to the algebraic Lyapunov equation
\begin{equation}
\label{ThetaLyap}
    \cA \Theta + \Theta \cA^{\rT} + \cB J \cB^{\rT} = 0.
\end{equation}
The left-hand side of (\ref{ThetaLyap}) is a real  antisymmetric matrix whose diagonal block, the upper off-diagonal block and the second diagonal block are computed as
\begin{align}
\label{11}
    A\Theta_1+\Theta_1 A^{\rT} + BJ_1B^{\rT} & =0,\\
\label{12}
    (\Theta_1 C^{\rT} + BJ_1 D^{\rT})e^{\rT}& =0, \\
\label{filterPR1}
    a\Theta_2 + \Theta_2 a^{\rT} +  e DJ_1 D^{\rT}e^{\rT} + bJ_2b^{\rT}&=0
\end{align}
in view of (\ref{cAB}), (\ref{J}), (\ref{cXCCR}).
The fulfillment of (\ref{11}), (\ref{12}) is guaranteed by the PR properties (\ref{plantPR1}), (\ref{plantPR2}) of the quantum plant for an arbitrary coherent quantum filter (\ref{xi_eta}), whereas (\ref{filterPR1}) and the equality
\begin{equation}
\label{filterPR2}
    \Theta_2c^{\rT} + bJ_2d^{\rT} =0
\end{equation}
describe PR conditions for the filter
which correspond to the preservation of the CCRs (\ref{xiCCR}) and $[\xi, \eta^{\rT}]=0$, respectively. The derivation of the PR conditions (\ref{filterPR1}), (\ref{filterPR2}) is similar to that of (\ref{plantPR1}), (\ref{plantPR2}) in (\ref{xx}), (\ref{xy}), except that the dynamic variables of the filter are driven not only by the quantum noise $\omega$, but also by the plant noise $w$ through the plant output $y$ according to  the QSDE  (\ref{closed}), which explains the presence of the additional term $e DJ_1D^{\rT}e^{\rT}$ in (\ref{filterPR1}). Since $\Theta_2$ is nonsingular, the general solution of
(\ref{filterPR1}), considered as a linear equation with respect to the matrix $a$, is described by
\begin{equation}
\label{agen}
    a
    =
    2\Theta_2 R
    -
    \frac{1}{2}
    (
        eDJ_1 D^{\rT} e^{\rT}
        +
        bJ_2 b^{\rT}
    )
    \Theta_2^{-1}.
\end{equation}
Here, $R$ is an arbitrary real symmetric matrix of order $q$ (the subspace of such matrices is denoted by $\mS_q$) which specifies  the quadratic Hamiltonian $\xi^{\rT}R\xi/2$ of an equivalent representation of
the PR filter as an open quantum harmonic oscillator.
%Therefore, under the assumption (\ref{Theta2nonsingular}), the state-space matrices of a PR coherent quantum filter are completely parameterized by the matrix triple $(b,e,R) \in \mR^{q\x m_2}\x \mR^{q\x p}\x \mS_q$.
Furthermore, (\ref{filterPR2}) allows the matrix $c$ to be expressed in terms of $b$ as
\begin{equation}
\label{c}
    c
    =
    -
    d
    J_2
    b ^{\rT}
    \Theta_2^{-1}.
\end{equation}
The coupling of the state-output matrix $c$  to the noise gain matrix $b$ makes the optimization of the coherent quantum filter (\ref{xi_eta}) qualitatively different from that of the classical controllers and filters, regardless of the performance criterion. Indeed, (\ref{c}) shows that the PR quantum filter requires an ``intake'' of the  additional quantum noise $\omega$ (through $b\ne 0$) in order to produce a useful output $\eta$ with a nonzero ``signal'' component
\begin{equation}
\label{zeta}
    \zeta:= c\xi.
\end{equation}
In their original form \cite{JNP_2008,NJP_2009,SP_2012}, % (see also Appendix~\ref{app:PR}),
the PR conditions also involve a specification of the  noise feedthrough matrices $D $, $d$ in (\ref{x_y}), (\ref{xi_eta}) as those formed from rows of orthogonal matrices, whereby
\begin{equation}
\label{DDdd}
    DD^{\rT} = I_p,
    \qquad
    dd^{\rT} = I_r.
\end{equation}
Therefore, $p\< m_1$, $r\< m_2$, and both $D$ and $d$ have full row rank. If $y$ in (\ref{x_y}) were a classical observation process,  the full row rank property of $D$ would correspond to nondegeneracy of the measurements. Also, since $d$ is of full row rank, the map $\mR^{q\x m_2}\ni b\mapsto c\in \mR^{r\x q}$, described by (\ref{c}), is surjective, so that the matrix $c$ can be assigned any value by an appropriate choice of $b$.
Although (\ref{DDdd}) will simplify algebraic manipulations, it is the rank properties of $D$, $d$  that are principal for what follows.

\section{Coherent Quantum Filtering Problem}\label{sec:problem}
%%%%%%%%%%%%%%%%%%%%%%%%%%%%%%%%%%%%%%%%%%%%%%%%%%%%%%%%%%%%%%%%%%%%%%%%%%%%%%%%

Consider the problem of constructing a PR  coherent quantum filter (\ref{xi_eta}) of fixed dimensions (and with a fixed noise feedthrough matrix $d$), described in Section~\ref{sec:filter}, so as to minimize a steady-state mean square discrepancy between the filter output variables and the state variables of a given PR quantum plant (\ref{x_y}) specified in Section~\ref{sec:plant}. More precisely, let $\cZ(t):= (\cZ_k(t))_{1\< k \< s}$ denote an $s$-dimensional quantum process defined by
\begin{equation}
\label{cZ}
    \cZ:= Fx - G \zeta
    =
    \cC
    \cX,
\end{equation}
where $F\in \mR^{s\x n}$, $G \in \mR^{s\x r}$ are given matrices, with $s\> r$ and $G$ having full column rank (the role  of this assumption is clarified later), and $\zeta$ is the signal part (\ref{zeta}) of the filter output $\eta$ in (\ref{xi_eta}), so that, in view of (\ref{cX}),
\begin{equation}
\label{cC}
    \cC:=
    \begin{bmatrix}
        F & -Gc
    \end{bmatrix}.
\end{equation}
The coherent quantum filtering (CQF) problem is formulated as the minimization of the quantity
\begin{align}
\nonumber
    \cE
    := &
    \lim_{t \to +\infty}
        \bE
        (
            \cZ ^{\rT} \cZ
        )\big/ 2
        =
    \bra
        \cC ^{\rT}\cC ,
        P
    \ket\big/ 2\\
\label{cE}
    & \longrightarrow
    \min,
    \quad{\rm subject\ to}\ (\ref{filterPR1}), (\ref{filterPR2}).
\end{align}
Here, the $1/2$ factor is introduced for further convenience, $\cZ^{\rT}\cZ= \sum_{k=1}^s \cZ_k^2$ is the sum of squared entries of $\cZ$, and
$\bE(\cdot)$ denotes the quantum expectation over the product state $\varpi\ox \upsilon$, where $\varpi$ is the initial quantum state of the plant-filter composite system on $\cH_1\ox \cH_2$, and $\upsilon$ is the vacuum state of the external fields. % in (\ref{ups}).
Also,   $\bra M,N\ket:= \Tr(M^*N)$ is the Frobenius inner product  of complex or real matrices, with $(\cdot)^*:= (\overline{(\cdot)})^{\rT}$ the complex conjugate transpose, and $P$ is the real part of the steady-state quantum covariance matrix of the vector $\cX$ in (\ref{cX}):
\begin{equation}
\label{KP}
    K:= \lim_{t\to +\infty}\bE (\cX\cX^{\rT}) = P+i\Theta,
    \qquad
    P:= \Re K,
\end{equation}
with $\lim_{t\to +\infty}\bE \cX = 0$, since the matrix $\cA$ in (\ref{cAB}) is Hurwitz. The latter condition is ensured by the Hurwitz property of the matrix $a$, since $\cA$ is block lower triangular and $A$ is Hurwitz. The matrix $K$ has the imaginary part $\Im K =  \Theta$ given by (\ref{cXCCR}) if the coherent quantum filter is also PR. In view of (\ref{closed}), (\ref{cWtable}), the matrix $P$ in (\ref{KP}) is a unique solution of the algebraic Lyapunov equation
\begin{equation}
\label{PLyap}
    \cA P + P \cA^{\rT} + \cB \cB^{\rT} = 0
\end{equation}
and coincides with the controllability Gramian \cite{KS_1972} of the pair $(\cA, \cB)$.
The minimum in the CQF problem (\ref{cE}) is taken over the quadruple $(a,b,c,e) \in
    \mR^{q\x q}
    \x
    \mR^{q\x m_2}
    \x
    \mR^{r\x q}
    \x
    \mR^{q\x p}
$
%\begin{equation}
%\label{u}
%    u:= (a,b,c,e)
%    \in
%    \mR^{q\x q}
%    \x
%    \mR^{q\x m_2}
%    \x
%    \mR^{r\x q}
%    \x
%    \mR^{q\x p}
%    =:
%    \mU,
%\end{equation}
of the state-space matrices
of the filter (\ref{xi_eta})  subject to the PR constraints (\ref{filterPR1}), (\ref{filterPR2}),
with a fixed noise feedthrough matrix $
    d\in \mR^{r\x m_2}
$ satisfying (\ref{DDdd}).
%The set $\mU$ in  (\ref{u}) is equipped with the structure of a Hilbert space with the direct sum inner product $\bra (a,b,c,e),(a',b',c',e') \ket := \bra a,a'\ket + \bra b,b'\ket + \bra c,c'\ket+ \bra e,e'\ket$.
%The full column rank condition
%\begin{equation}
%\label{Gfullcolumnrank}
%    G^{\rT}G\succ 0
%\end{equation}
%on the matrix $G$ ensures strict convexity of the function $\cE$ in (\ref{cE}) with respect to the matrix $c$. %
%For what follows, we assume that
%$    r\< s\< n
%$.
In particular, if $F=G=I_n$, with  $r=s=n$, the CQF problem (\ref{cE}) consists in approximating  the plant state vector $x$ by the signal part $\zeta$ of the filter output $\eta$ from (\ref{zeta}) so as to minimize the ``estimation error'' $x-\zeta$ in the mean square sense.
\section{Qualitative Dependence on Filter Matrices}\label{sec:dep}
%%%%%%%%%%%%%%%%%%%%%%%%%%%%%%%%%%%%%%%%%%%%%%%%%%%%%%%%%%%%%%%%%%%%%%%%%%%%%%%

The performance index in the CQF problem (\ref{cE}) is a composite function
$
    (a,b,c,e)\mapsto (\cA, \cB, \cC) \mapsto \cE
$, where the triple
$    (\cA, \cB, \cC)
    \in
    \mR^{(n+q)\x (n+q)}
    \x
    \mR^{(n+q)\x m}
    \x
    \mR^{s\x (n+q)}
$
of matrices from (\ref{cAB}), (\ref{cC}) depends affinely on the matrix quadruple $(a,b,c,e)$ (with both being regarded as elements of appropriate direct sum Hilbert spaces). The smoothness %(moreover, infinite differentiability)
of $\cE$ with respect to  $\cA$, $\cB$, $\cC$ (regardless of the specific structure of these matrices) is ensured by the smooth dependence of the controllability Gramian
\begin{equation}
\label{PLam}
    P
    =
    \Lambda_{\cA}(\cB \cB^{\rT})
\end{equation}
from (\ref{PLyap}) on $\cA$, $\cB$ over the open subset of Hurwitz matrices $\cA$. Here, $\Lambda_{\alpha}$ denotes the inverse Lyapunov operator, that is, a particular case of the inverse Sylvester operator $\Sigma_{\alpha, \beta}$,
which is associated with Hurwitz matrices  $\alpha$, $\beta$  and maps an appropriately dimensioned  matrix $X$ to the  unique solution $Y:= \Sigma_{\alpha,\beta}(X)$ of the algebraic Sylvester  equation $\alpha Y + Y\beta + X=0$:
\begin{equation}
\label{Sigma}
    \Lambda_{\alpha} := \Sigma_{\alpha, \alpha^{\rT}},
    \qquad
    \Sigma_{\alpha,\beta}(X)
    :=
    \int_{0}^{+\infty}
    \re^{\alpha t}
    X
    \re^{\beta t}
    \rd t.
\end{equation}
%so that $\Lambda_{\alpha} = \Sigma_{\alpha, \alpha^{\rT}}$.
For what follows,
the controllability Gramian $P$ in (\ref{PLam}) (and other related matrices of order $n+q$) is split into blocks $P_{11}\in \mS_n$, $P_{22}\in\mS_q$, $P_{12}=P_{21}^{\rT} \in \mR^{n\x q}$ and the corresponding block-rows $P_{j\bullet}$ and block-columns $P_{\bullet k}$ in accordance with their association with the plant and filter variables in (\ref{cX}) as
\begin{equation}
\label{Pblocks}
    P
    :=
    {\small\begin{bmatrix}
                P_{11} & P_{12}\\
                P_{21} & P_{22}
    \end{bmatrix}}
    =
    {\small\begin{bmatrix}
                P_{1\bullet}\\
                P_{2\bullet}
    \end{bmatrix}}
    =
    \begin{bmatrix}
                P_{\bullet 1} & P_{\bullet 2}
    \end{bmatrix}.
\end{equation}
The Lyapunov equation (\ref{PLyap}), whose left-hand side is a symmetric matrix with a similar block partitioning, can be written in terms of (\ref{Pblocks}) as
\begin{align}
\label{PLyap11}
    AP_{11}+P_{11}A^{\rT} + BB^{\rT} & =0,\\
\label{PLyap12}
    AP_{12} + P_{12}a^{\rT} + P_{11} C^{\rT}e^{\rT} + BD^{\rT}e^{\rT}& =0, \\
\label{PLyap22}
    aP_{22} + P_{22}a^{\rT} + eCP_{12} + P_{21} C^{\rT}e^{\rT} + e %DD^{\rT}
    e^{\rT} + bb^{\rT}&=0,
\end{align}
where (\ref{cAB}), (\ref{DDdd}) are used. The inverse Lyapunov and Sylvester operators (\ref{Sigma})  allow the solution of (\ref{PLyap11})--(\ref{PLyap22}) to be represented as
\begin{align}
\label{P11}
    P_{11} & = \Lambda_A (BB^{\rT}),\\
\label{P12}
    P_{12} & = \Sigma_{A,a^{\rT}}((P_{11} C^{\rT} + BD^{\rT})e^{\rT}), \\
\label{P22}
    P_{22} & = \Lambda_a(eCP_{12} + P_{12}^{\rT} C^{\rT}e^{\rT} + e%DD^{\rT}
    e^{\rT} + bb^{\rT}).
\end{align}
Since the matrix $P_{11}$ in (\ref{P11}) is specified completely by the quantum plant (\ref{x_y}) and does not depend on the coherent quantum filter,  the matrix $P_{11} C^{\rT} + BD^{\rT}$ in (\ref{P12}) is also constant. Therefore, for any given Hurwitz matrix $a$ of the filter (\ref{xi_eta}), the matrix $P_{12}$ is a  linear function of $e$. Hence, $P_{22}$ in (\ref{P22}) is a homogeneous quadratic polynomial
of the filter matrices $b$, $e$ from (\ref{xi_eta}), whose coefficients depend on $a$ and which does not contain  the $b_{jk}e_{\ell z}$ cross-terms with mixed entries of $b$, $e$.
% which does not involve the $b_{jk}e_{\ell z}$ cross-terms:
%\begin{equation}
%\label{P22pol}
%    P_{22} =
%    \sum_{j,\ell=1}^{q}
%    \bigg(
%        \sum_{k,s=1}^{m_2}
%        \Pi_{jk\ell s}
%        b_{jk}b_{\ell s}
%        +
%        \sum_{v,z=1}^{p}
%        \mho_{jv\ell z}
%        e_{jv}e_{\ell z}
%    \bigg).
%\end{equation}
%The matrix-valued coefficients
%$\Pi_{jk\ell s} \in \mS_q$ and   $\mho_{jv\ell z} \in \mS_q$ of the polynomial depend on $a$.
 The performance index $\cE$ in the CQF problem (\ref{cE}) is representable in terms of the block partitioning (\ref{Pblocks}) as
\begin{equation}
\label{cEblocks}
    \cE =
    \bra F^{\rT}F, P_{11}\ket/2
    -
    \bra F^{\rT}Gc, P_{12}\ket
    +
    \bra c^{\rT}G^{\rT}Gc, P_{22}\ket    /2,
\end{equation}
where (\ref{cC}) is used together with the symmetry of the matrices $\cC^{\rT}\cC$ and $P$. In combination with the quadratic dependence of $P_{22}$ on $b$, $e$,
and the linear dependence of $P_{12}$ on $e$ discussed above, (\ref{cEblocks}) implies that  $\cE$ is quadratic in $b$, $c$, $e$ for any fixed Hurwitz matrix $a$. Such dependence of the quadratic performance index $\cE$ on the filter matrices also holds in the classical filtering problem. However, in contrast to its classical predecessor, the CQF problem (\ref{cE}) is constrained by the PR conditions (\ref{filterPR1}), (\ref{filterPR2}) which couple the skew-Hamiltonian part of the matrix $a$ in (\ref{agen}) to $b$, $e$ and the matrix $c$ in (\ref{c}) to $b$.

%%%%%%%%%%%%%%%%%%%%%%%%%%%%%%%%%%%%%%%%%%%%%%%%%%%%%%%%%%%%%%%%%%%%%%%%%%%%%%%
\section{Conditions of Stationarity}\label{sec:diff}
%%%%%%%%%%%%%%%%%%%%%%%%%%%%%%%%%%%%%%%%%%%%%%%%%%%%%%%%%%%%%%%%%%%%%%%%%%%%%%%

If an optimal filter exists for the CQF problem (\ref{cE}), such a filter is among stationary points of
the Lagrange function
\begin{align}
\nonumber
    \cL
    := &
    \cE +\bra \Xi, a\Theta_2 + \Theta_2a^{\rT} + eDJ_1 D^{\rT} e^{\rT} + bJ_2 b^{\rT}\ket/ 2\\
\nonumber
     & + \bra \Gamma, \Theta_2c^{\rT} + bJ_2d^{\rT}\ket\\
\nonumber
    = &
    \cE - \bra \Xi \Theta_2, a\ket
    +
    \bra \Xi, eDJ_1 D^{\rT} e^{\rT} + bJ_2 b^{\rT}\ket/ 2\\
\label{cL}
    &
    -
    \bra \Gamma d J_2, b\ket + \bra \Gamma^{\rT} \Theta_2, c\ket.
\end{align}
Here, the $1/2$ factor is introduced for further convenience, and $\Xi \in \mA_q$, $\Gamma \in \mR^{q\x r}$ are Lagrange multipliers associated with the PR constraints (\ref{filterPR1}) (whose left-hand side is a real antisymmetric matrix) and  (\ref{filterPR2}), respectively.  We will be concerned with a quadruple $(a,b,c,e)$ %(\ref{u})
of \emph{unconstrained}  state-space matrices of the coherent quantum filter (\ref{xi_eta})
%(with the noise feedthrough matrix $d$ being fixed and satisfying (\ref{DDdd}))
which is a stationary point of the Lagrange function  $\cL$ in (\ref{cL}) for given matrices $\Xi$, $\Gamma$, with $a$ Hurwitz. The Lagrange multipliers $\Xi$, $\Gamma$ are to be found so as to ensure that the filter (which depends parametrically  on $\Xi$, $\Gamma$) satisfies the PR conditions. In order to find stationary points of the Lagrange function $\cL$ in (\ref{cL}),  we will compute its Frechet derivatives by using the chain rule and the following lemma \cite[Lemma 2]{VP_2011a} based on algebraic techniques from
\cite{BH_1998,SIG_1998}.

%==============================================================================
\begin{lem}
\label{lem:dEdABC}
%Suppose the matrix $\cA$ in (\ref{cAB}) is Hurwitz. Then
The Frechet derivatives of the function $ \cE $ in (\ref{cE})   with respect to the matrices $\cA$, $\cB$, $\cC$ from (\ref{cAB}), (\ref{cC}), with $\cA$ Hurwitz, are computed as
\begin{equation}
\label{dEdABC}
        \d_{\cA}\cE = QP=: H,
        \qquad
        \d_{\cB}\cE = Q\cB,
        \qquad
        \d_{\cC}\cE = \cC P,
\end{equation}
where $P$ is the controllability Gramian from (\ref{PLyap}), and $Q:= \Lambda_{\cA^{\rT}}(\cC^{\rT}\cC)$ is the observability Gramian of the pair $(\cA, \cC)$ satisfying the algebraic Lyapunov equation
\begin{equation}
\label{QLyap}
    \cA^{\rT} Q + Q \cA + \cC^{\rT} \cC = 0.
\end{equation}
\end{lem}\hfill$\square$
%==============================================================================

The matrix $H$ in (\ref{dEdABC}) will be referred to as the \emph{Hankelian} of the triple $(\cA,\cB,\cC)$  since the  spectrum of $H$ is formed by the squared Hankel singular values \cite{KS_1972,SIG_1998} of an appropriate linear time-invariant system.
A block-wise form of (\ref{QLyap}) is given by
\begin{align}
\label{QLyap11}
    A^{\rT}Q_{11}+Q_{11}A + C^{\rT} e^{\rT} Q_{21} + Q_{12}eC + F^{\rT} F& =0,\\
\label{QLyap21}
    a^{\rT}Q_{21} + Q_{21}A + Q_{22} eC - c^{\rT} G^{\rT}F& =0, \\
\label{QLyap22}
    a^{\rT}Q_{22} + Q_{22}a + c^{\rT} G^{\rT}Gc&=0,
\end{align}
which is similar to (\ref{PLyap11})--(\ref{PLyap22}) except that the lower off-diagonal block is considered instead of the upper one. The solution of (\ref{QLyap11})--(\ref{QLyap22}) is found by using the inverse Lyapunov and Sylvester operators (\ref{Sigma}) as
\begin{align}
\label{Q11}
    Q_{11} & = \Lambda_{A^{\rT}}(C^{\rT} e^{\rT} Q_{21} + Q_{21}^{\rT}eC + F^{\rT} F),\\
\label{Q21}
    Q_{21} & = \Sigma_{a^{\rT},A}(Q_{22} eC - c^{\rT} G^{\rT}F), \\
\label{Q22}
    Q_{22} & = \Lambda_{a^{\rT}}(c^{\rT} G^{\rT}Gc),
\end{align}
which corresponds to (\ref{P11})--(\ref{P22}).
The block $Q_{22}$ in (\ref{Q22}) is computed first and is then substituted into (\ref{Q21}) in order to find $Q_{21}$, while $Q_{11}$ in (\ref{Q11}) is irrelevant for further discussions.

%In accordance with the partitioning of $\cX$ into the plant and filter variables in (\ref{cX}), the Gramians $P$, $Q$ and the Hankelian $H$ (and other related matrices of order $n+q$ such, for example, as $Q\cA P$, $\cC^{\rT}\cC$, $\cB\cB^{\rT}$) are split into four blocks $P_{jk}$, $Q_{jk}$ and $H_{jk}$, $1\<j,k\< 2$, respectively, as
%\begin{align}
%\nonumber
%    H
%    := &
%    {\small\begin{bmatrix}
%                H_{11} & H_{12}\\
%                H_{21} & H_{22}
%    \end{bmatrix}}
%    =
%    {\small\begin{bmatrix}
%                H_{1\bullet}\\
%                H_{2\bullet}
%    \end{bmatrix}}
%    =
%    {\small\begin{bmatrix}
%                H_{\bullet 1} & H_{\bullet 2}
%    \end{bmatrix}},\\
%\label{Hblocks}
%    H_{j\bullet}
%    := &
%    {\small\begin{bmatrix}
%                H_{j1}&
%                H_{j2}
%    \end{bmatrix}},
%    \qquad
%    H_{\bullet k}
%    :=
%    {\small\begin{bmatrix}
%                H_{1k}\\
%                H_{2k}
%    \end{bmatrix}},
%\end{align}
%where $P_{j\bullet}$, $Q_{j\bullet}$, $H_{j\bullet}$ denote the appropriate block-rows, and  $P_{\bullet k}$, $Q_{\bullet k}$, $H_{\bullet k}$ are the block-columns  of the matrices. The dimensions of the blocks are recovered from their association with the plant and filter state variables: $H_{11}\in \mR^{n\x n}$, $H_{21}\in \mR^{q\x n}$,  $H_{12}\in \mR^{n\x q}$,  $H_{22}\in \mR^{q\x q}$.

%==============================================================================
\begin{lem}
\label{lem:dEdabce}
%Suppose the matrix $\cA$ in (\ref{cAB}) is Hurwitz. Then t
The Frechet derivatives of the Lagrange function $\cL$ from (\ref{cL}) with respect to the state-space matrices $a$, $b$, $c$, $e$ of the quantum filter (\ref{xi_eta}), with $a$ Hurwitz,  are computed as
\begin{align}
\label{dLda}
    \d_a \cL
    & =
    H_{22} - \Xi \Theta_2,\\
\label{dLdb}
    \d_b \cL
    & =
    Q_{22} b %+ \Theta_2^{-1} P_{2\bullet}\cC^{\rT} Gd J_2
    -\Xi b J_2 - \Gamma d J_2,\\
\label{dLdc}
    \d_c \cL
    & =
    -G^{\rT} F P_{12} + G^{\rT}G c P_{22}
    + \Gamma^{\rT}\Theta_2,\\
\label{dLde}
    \d_e \cL
    & =
    H_{21} C^{\rT} + Q_{21} BD^{\rT} + Q_{22}e%DD^{\rT}
    - \Xi eDJ_1D^{\rT},
\end{align}
where use is made of the block partitioning  of the controllability and observability Gramians $P$, $Q$ and the Hankelian $H$ according to (\ref{Pblocks}).\hfill$\square$
\end{lem}
%==================================================================================================

The proof of Lemma~\ref{lem:dEdabce} is given in Appendix~\ref{app:proof_lem:dEdabce}.
The following conditions of stationarity of the Lagrange function $\cL$ are obtained by equating its Frechet derivatives to zero.

%==================================================================================================
\begin{thm}
\label{thm:stat}
%Suppose the matrix $\cA$ in (\ref{cAB}) is Hurwitz. Then
The coherent quantum filter (\ref{xi_eta}), with $a$ Hurwitz, is a stationary point of the CQF problem (\ref{cE}) if and only if there exist Lagrange multipliers $\Xi\in \mA_q$, $\Gamma\in \mR^{q\x r}$ in (\ref{cL}) such that the equalities
\begin{align}
\label{dLda0}
    H_{22} & = \Xi \Theta_2,\\
\label{dLdb0}
    Q_{22} b
    &=\Xi b J_2 + \Gamma d J_2,\\
\label{dLdc0}
    G^{\rT}G c P_{22}
    -
    G^{\rT} F P_{12}
    &=- \Gamma^{\rT}\Theta_2,\\
\label{dLde0}
    H_{21} C^{\rT} + Q_{21} BD^{\rT} + Q_{22}e%DD^{\rT}
    &= \Xi eDJ_1D^{\rT}
\end{align}
are satisfied together with the PR conditions (\ref{filterPR1}), (\ref{filterPR2}).\hfill$\square$
\end{thm}
%==================================================================================================

Together with the Lyapunov equations (\ref{PLyap}), (\ref{QLyap}) and the PR conditions (\ref{filterPR1}), (\ref{filterPR2}), the relations (\ref{dLda0})--(\ref{dLde0}) form a complete set of algebraic equations for finding an optimal filter  among stationary points $(a,b,c,e,\Xi,\Gamma)$ of the Lagrange function in (\ref{cL}). Since (\ref{dLdb0}), (\ref{dLdc0}) can, in principle,  be solved for $b$, $c$, the Lagrange multiplier $\Gamma$ can be found so as to satisfy the PR condition (\ref{filterPR2}). Also, (\ref{dLde0}) can be solved for $e$, as discussed below. In view of  (\ref{dLda0}), which describes the stationarity of the Lagrange function with respect to $a$,  and the antisymmetry of $\Xi$,  the matrix $H_{22} = \Theta_2^{-1}\Theta_2 \Xi \Theta_2$ is skew-Hamiltonian in the sense of the symplectic structure specified by $\Theta_2^{-1}$. Note that (\ref{dLda0}) is implicit in $a$.  In the next section, we will obtain a more explicit equation and discuss its solvability with respect to $a$ along with that of (\ref{dLdb0})--(\ref{dLde0}) for $b$, $c$, $e$.

\section{Combining the Stationarity and Lyapunov Equations}\label{sec:stat}
%%%%%%%%%%%%%%%%%%%%%%%%%%%%%%%%%%%%%%%%%%%%%%%%%%%%%%%%%%%%%%%%%%%%%%%%%%%%%%%

We will need two identities for the Gramians $P$, $Q$ from (\ref{PLyap}), (\ref{QLyap}) and the Hankelian $H$ from (\ref{dEdABC}) which hold  regardless of whether the QSDEs (\ref{x_y}), (\ref{xi_eta}) are PR. To this end, we introduce a matrix
\begin{equation}
\label{Q2AP2}
    \Ups :=Q_{2\bullet} \cA P_{\bullet 2}
    =
    Q_{21} A P_{12} + Q_{22} eC P_{12} + Q_{22} aP_{22},
\end{equation}
where use is made of (\ref{cAB}) along with the block partitioning of $P$, $Q$, $H$ in accordance with (\ref{Pblocks}).

%==================================================================================================
\begin{lem}
\label{lem:QAP}
%Suppose the matrix $\cA$ in (\ref{cAB}) is Hurwitz.
The matrix $\Ups $, defined by (\ref{Q2AP2}), with $\cA$ Hurwitz,  satisfies
\begin{align}
\nonumber
    \Ups
    +
    (H_{21} C^{\rT} + Q_{21} BD^{\rT} + Q_{22} e%DD^{\rT}
    ) e^{\rT} &\\
\label{QAP1}
     + H_{22}a^{\rT} + Q_{22} bb^{\rT} &=0,\\
\label{QAP2}
    \Ups
    +
    a^{\rT} H_{22} + c^{\rT} G^{\rT} (Gc P_{22}-FP_{12}) &= 0.
\end{align}
\end{lem}\hfill$\square$
%==================================================================================================

The proof of Lemma~\ref{lem:QAP} is given in Appendix~\ref{app:proof_lem:QAP}.
We will now combine the general identities (\ref{QAP1}), (\ref{QAP2}) with the equations (\ref{dLda0})--(\ref{dLde0}) of stationarity   of the Lagrange function $\cL$ from Theorem~\ref{thm:stat} and the PR conditions (\ref{filterPR1}), (\ref{filterPR2}).

%==================================================================================================
\begin{lem}
\label{lem:sym}
%Suppose the matrix $\cA$ in (\ref{cAB}) is Hurwitz, and (\ref{Theta2nonsingular}) holds.
If the PR coherent quantum filter is a stationary point of the  Lagrange function in (\ref{cL}), with $a$ Hurwitz, then the matrix $\Xi a + \Gamma c$ is symmetric, and the matrix $\Ups $ from (\ref{Q2AP2}) satisfies
\begin{equation}
\label{QAP1eq}
    \Ups  = (\Xi a + \Gamma c)\Theta_2.
\end{equation}
\end{lem}\hfill$\square$
%==================================================================================================

The proof of Lemma~\ref{lem:sym} is provided in Appendix~\ref{app:proof_lem:sym}. The lemma will be used to obtain a more explicit equation for $a$ than (\ref{dLda0}).
Now, suppose the ``filter blocks''  of the controllability and observability Gramians $P$, $Q$ are both nonsingular:
 \begin{equation}
 \label{PQpos}
    P_{22}\succ 0,
    \qquad
    Q_{22}
    \succ 0.
 \end{equation}
The positive semi-definiteness  $P_{22}\succcurlyeq 0$, $Q_{22}\succcurlyeq 0$ is inherited from $P$, $Q$. By the general form of Heisenberg's uncertainty principle \cite{H_2001}, the steady-state quantum covariance matrix of the filter state variables satisfies
  \begin{equation}
  \label{S22pos}
    P_{22} + i\Theta_2
    =
    \lim_{t\to +\infty}\bE(\xi\xi^{\rT})
    \succcurlyeq
    0,
  \end{equation}
  which is a stronger property than $P_{22}\succcurlyeq 0$. The assumptions (\ref{PQpos})
  %are of technical nature
  %and
  enable the following matrices to be defined:
  \begin{align}
\label{LM}
    L
    & :=
    Q_{22}^{-1}Q_{21},
    \qquad
    M
    :=
    P_{12}
    P_{22}^{-1},\\
\label{NS}
    N
     & :=
    Q_{22}^{-1}\Xi,
    \qquad\quad\,
    S
    :=
    \Theta_2 P_{22}^{-1},\\
\label{TU}
    T
    &:=
    Q_{22}^{-1} \Gamma,
    \qquad\quad
    U
    :=
    Q_{22} \Theta_2 P_{22}^{-1}.
\end{align}
Here, only $N$, $T$ involve the Lagrange multipliers $\Xi$, $\Gamma$ from (\ref{cL}), while $L$, $M$, $S$, $U$ are completely specified by $\Theta_2$ and the appropriate blocks of the Gramians $P$, $Q$, with neither $P_{11}$ nor $Q_{11}$ being involved. The eigenvalues of the matrix $S$ in (\ref{NS}) are purely imaginary and symmetric about the origin, with the spectral radius satisfying
\begin{equation}
\label{brS}
    \br(S)\< 1
\end{equation}
in view of (\ref{S22pos}).
%Under the assumption  $P_{22}\succ 0$ in (\ref{PQpos}),  the positive semi-definiteness in (\ref{S22pos}) implies that the spectral radius of the matrix $S$ from (\ref{NS}), whose eigenvalues are purely imaginary and symmetric about the origin, satisfies $\br(S)\< 1$. Moreover, the positive semi-definiteness
%\begin{equation}
%\label{PPPP}
%    \begin{bmatrix}
%        P_{11} + i\Theta_1 & P_{12}\\
%        P_{21} & P_{22} + i\Theta_2
%    \end{bmatrix}
%    =
%    \lim_{t\to +\infty} \bE(\cX\cX^{\rT})
%    \succcurlyeq 0
%\end{equation}
%of the steady-state quantum covariance matrix  of the plant-filter state vector $\cX$ from (\ref{cX}), considered under an additional assumption that $P_{22}+i\Theta_2\succ 0$ (that is, $\br(S)<1$), is equivalent to
%\begin{align*}
%    P_{12}(P_{22} + i\Theta_2)^{-1}P_{21}
%    =&
%    M(I_q+S^2)^{-1} (I_q - iS) P_{21}\\
%    \preccurlyeq&  P_{11} + i \Theta_1,
%\end{align*}
%which involves the matrix $M$ from (\ref{LM}). Here, use is made of the identity
%$
%    (\alpha +i\beta)^{-1}
%    =
%    \alpha^{-1}(I + \gamma^2)^{-1}    (I - i \gamma)
%$
%for square matrices $\alpha$, $\beta$ of same order, with     $\gamma:= \beta\alpha^{-1}$,
%provided $\alpha$ and $I+\gamma^2$ are both nonsingular.
The matrix ``ratios'' from (\ref{LM})--(\ref{TU}) allow the equations of Theorem~\ref{thm:stat} to be made %represented in a form which is
more explicit in $a$, $b$, $c$, $e$.

%%%%%%%%%%%%%%%%%%%%%%%%%%%%%%%%%%%%%%%%%%%%%%%%%%%%%%%%%%%%%%%%%%%%%%%%%%%%%%%%%%%%%%%%%%%%%%%%%%%
\begin{lem}
\label{lem:rat}
Under the assumptions (\ref{PQpos}), the stationarity equations (\ref{dLda0})--(\ref{dLde0}) for the CQF problem (\ref{cE}) can be represented  in terms of the matrices $L$, $M$, $N$, $S$, $T$, $U$ from (\ref{LM})--(\ref{TU}) as
\begin{align}
\label{da0}
    LM & = NS-I_q,\\
\label{db0}
    b & = NbJ_2 + TdJ_2,\\
\label{dc0}
    c & = (G^{\rT}G)^{-1} (G^{\rT} F M - T^{\rT} U),\\
\label{de0}
    e & = Ne\Delta -(L(P_{11}C^{\rT} + BD^{\rT}) + P_{21}C^{\rT}),%(DD^{\rT})^{-1},
\end{align}
where
\begin{equation}
\label{Delta}
    \Delta :=
    DJ_1D^{\rT}.
%    (DD^{\rT})^{-1}.
\end{equation}
Furthermore, the matrix $a$ of such a filter satisfies the relation
\begin{equation}
\label{a}
    a
    =
    (Na + Tc)S
    -(LA+eC)M.
\end{equation}
\hfill$\square$
\end{lem}
%%%%%%%%%%%%%%%%%%%%%%%%%%%%%%%%%%%%%%%%%%%%%%%%%%%%%%%%%%%%%%%%%%%%%%%%%%%%%%%%%%%%%%%%%%%%%%%%%%%
The proof of Lemma~\ref{lem:rat} is given in Appendix~\ref{app:proof_lem:rat}. Note that (\ref{dc0}) employs the full column rank assumption on the matrix $G$ in (\ref{cZ}) to ensure the invertibility of $G^{\rT}G$.  While (\ref{dc0}) is already solved for $c$, we will show how the stationarity equations (\ref{a}), (\ref{db0}), (\ref{de0}) can be solved for the other filter matrices $a$, $b$, $e$. To this end, let $\[[[\alpha, \beta\]]]: X \mapsto \alpha X \beta$ denote the  linear operator of the left and right multiplication of an appropriately dimensioned matrix $X$ by given real matrices $\alpha$, $\beta$, respectively.
More generally \cite[Section 7]{VP_2011a}, for any positive integer $g$ and compatibly dimensioned matrices $\alpha_1, \ldots, \alpha_g$ and $\beta_1, \ldots, \beta_g$, a \emph{special linear operator of grade} $g$ is defined by
\begin{equation}
\label{specoper}
    \[[[
        \alpha_1,\beta_1
        \mid
        \ldots
        \mid
        \alpha_g, \beta_g
    \]]]
    :=
    \sum_{k=1}^g
    \[[[
        \alpha_k,  \beta_k
    \]]],
\end{equation}
where the matrix pairs are separated by ``$\mid$''s. If, in each of the pairs, the matrices $\alpha_k$, $\beta_k$ are either both symmetric or both antisymmetric, then the operator (\ref{specoper}) is self-adjoint with respect to the Frobenius inner product of matrices. Now, the solvability of (\ref{a}) with respect to $a$ depends on whether the grade two special linear operator
\begin{equation}
\label{opa}
    \[[[ I_q,I_q \mid -N, S\]]]
    =
    \[[[Q_{22}^{-1}, P_{22}^{-1}\]]]
    \circ
    \[[[Q_{22}, P_{22}\mid -\Xi, \Theta_2\]]]
\end{equation}
(with  $\circ$ denoting the composition) is invertible on $\mR^{q\x q}$.
%Here, the invertibility $\[[[Q_{22}^{-1}, P_{22}^{-1}\]]] = \[[[Q_{22}, P_{22}\]]]^{-1}$ is secured by (\ref{PQpos}).
In the case of invertibility, (\ref{a}) takes the form
\begin{equation}
\label{aa}
    a
    =
    \[[[
        I_q,I_q
        \mid
        -N, S
    \]]]^{-1}(TcS-(LA+eC)M).
\end{equation}
In a similar vein, (\ref{db0}), (\ref{de0}) can be solved for $b$, $e$, respectively, as
\begin{align}
\label{b}
    b
    & =
    \[[[
        I_q,I_{m_2}
        \mid
        -N, J_2
    \]]]^{-1}(TdJ_2),\\
\label{e}
    e
    & =
    -
    \[[[
        I_q,I_p
        \mid
        -N, \Delta
    \]]]^{-1}(L(P_{11}C^{\rT} + BD^{\rT}) + P_{21}C^{\rT}),
\end{align}
provided the following special linear operators of grade two are invertible:
\begin{align}
\label{opb}
    \[[[
        I_q,I_{m_2}
        \mid
        -N, J_2
    \]]]
    &=
    \[[[
        Q_{22}^{-1}, I_{m_2}
    \]]]
    \circ
    \[[[
        Q_{22}, I_{m_2}
        \mid
        -\Xi, J_2
    \]]],\\
\label{ope}
    \[[[
        I_q,I_p
        \mid
        -N, \Delta
    \]]]
    &=
    \[[[
        Q_{22}^{-1}, I_p
    \]]]
    \circ
    \[[[
        Q_{22}, I_p
        \mid
        -\Xi, \Delta
    \]]].
\end{align}
By combining the spectral property (\ref{brS}) of the matrix $S$ from (\ref{NS}) with similar properties
\begin{equation}
\label{rr}
    \br(J_2) = 1,
    \qquad
    \br(\Delta) \< 1
\end{equation}
of the matrices $J_2$ in (\ref{J2}) and $\Delta$ in (\ref{Delta}), and applying Lemma~\ref{lem:two} of Appendix~\ref{app:two} to the operators in (\ref{opa}), (\ref{opb}), (\ref{ope}), it follows that the condition
\begin{equation}
\label{brN}
    \br(N)< 1
\end{equation}
is sufficient for the invertibility of all three operators. Note that, similarly to \cite[Proof of Lemma 5]{VP_2011a}, the second of the spectral relations (\ref{rr}) follows from (\ref{canJ1}) and (\ref{DDdd}) which imply that the complex Hermitian matrix $i\Delta=iDJ_1D^{\rT}$ satisfies $-I_p\preccurlyeq i\Delta \preccurlyeq I_p$. Under the assumption $Q_{22}\succ 0$, the condition (\ref{brN}) is equivalent to the strict convexity of the Lagrange function $\cL$ in (\ref{cL}) with respect to $b$. More precisely, $\cL$ inherits quadratic dependence on $b$ from $\cE$, as discussed in Section~\ref{sec:dep}, and, in view of (\ref{dLdb}), the    second order Frechet derivative
$
    \d_b^2 \cL
    =
    \[[[
        Q_{22}, I_{m_2}
        \mid
        -\Xi, J_2
    \]]]
$
is a grade two special self-adjoint operator on the Hilbert space $\mR^{q\x m_2}$ whose positive definiteness is indeed equivalent to (\ref{brN}) by Lemma~\ref{lem:two}. Here, we have again used the spectral property of the matrix $J_2$ from (\ref{rr}).

%%%%%%%%%%%%%%%%%%%%%%%%%%%%%%%%%%%%%%%%%%%%%%%%%%%%%%%%%%%%%%%%%%%%%%%%%%%%%%%
\section{An Iterative Algorithm Outline}\label{sec:num}
%%%%%%%%%%%%%%%%%%%%%%%%%%%%%%%%%%%%%%%%%%%%%%%%%%%%%%%%%%%%%%%%%%%%%%%%%%%%%%%

A reasoning, similar to that in the previous section, shows
that, under the assumptions (\ref{PQpos}) and (\ref{brN}),  the matrices $b$, $c$ in (\ref{db0}), (\ref{dc0}) satisfy the PR constraint (\ref{filterPR2}) if and only if the matrix $T$ in (\ref{TU}) is related to $b$ by
\begin{equation}
\label{TT}
    T
    =
    U^{-\rT}(M^{\rT}F^{\rT}G + \Theta_2^{-1}bJ_2 d^{\rT} G^{\rT}G),
\end{equation}
where $(\cdot)^{-\rT}:= ((\cdot)^{-1})^{\rT}$ is the composition of the matrix inverse and transpose. This  allows $T$ to be eliminated from (\ref{b}) as
\begin{align}
\nonumber
    b
    = &
    \[[[
        I_q,I_{m_2}
         \mid
        -N, J_2\\
\label{bb}
        & \mid
        -U^{-\rT} \Theta_2^{-1},
        J_2 d^{\rT} G^{\rT}G dJ_2
    \]]]^{-1}
    (U^{-\rT}M^{\rT}F^{\rT}GdJ_2),
\end{align}
which involves an invertible special operator of grade three.  The matrix $b$ from (\ref{bb}) can be  substituted into (\ref{c}) in order to find $c$. In an iterative algorithm for numerical computation of an optimal quantum filter in the CQF problem (\ref{cE}), the relations (\ref{bb}), (\ref{TT}), (\ref{c}) can be employed to update the matrices $b$, $T$, $c$ for given matrices $M$, $N$, $U$ from (\ref{LM})--(\ref{TU}). In a similar vein, (\ref{aa}), (\ref{e}) can be used for updating the matrices $a$, $e$  for given $L$, $M$, $N$, $S$, $T$ and $P_{12}$. This is accompanied by updating the matrices $L$, $M$, $S$, $U$ according to (\ref{LM})--(\ref{TU}) in terms of the blocks of the Gramians $P$, $Q$ which are computed for given filter matrices $a$, $b$, $c$, $e$ as described by (\ref{P12}), (\ref{P22}) and (\ref{Q21}), (\ref{Q22}). The algorithm loop can be closed by computing the block $H_{22}$ of the Hankelian $H$ in (\ref{dEdABC}) and updating the Lagrange multiplier $\Xi$ in (\ref{NS}) as
$
    \Xi
    :=
    (H_{22}\Theta_2^{-1} + \Theta_2^{-1} H_{22}^{\rT})/2
$. In the case $n=q$, the algorithm can be initialized with the state-space matrices of a classical Kalman filter from the next section.

%Furthermore, (\ref{QAP1eq}) of Lemma~\ref{lem:sym} takes the form
%The spectrum of the matrix $\Delta$ is purely imaginary and symmetric about the origin and satisfies $\br(\Delta)\< 1$;

%$$
%    Q_{22}((LA + eC) M + a)P_{22} = (\Xi a + \Gamma c)\Theta_2
%$$

%\begin{equation}
%\label{Gamma}
%    \Gamma
%    =
%    -\Theta_2^{-1}P_{21}F^{\rT}G - \Theta_2^{-1} P_{22}\Theta_2^{-1} b J_2 d^{\rT}G^{\rT}G
%\end{equation}

%%%%%%%%%%%%%%%%%%%%%%%%%%%%%%%%%%%%%%%%%%%%%%%%%%%%%%%%%%%%%%%%%%%%%%%%%%%%%%%
\section{Reduction to the Classical Kalman Filter}\label{sec:class}
%%%%%%%%%%%%%%%%%%%%%%%%%%%%%%%%%%%%%%%%%%%%%%%%%%%%%%%%%%%%%%%%%%%%%%%%%%%%%%%

If the PR constraints (\ref{filterPR1}), (\ref{filterPR2}) are made inactive  by letting
$\Sigma=0$, $\Gamma=0$ in (\ref{cL}) (so that  the CCRs become irrelevant), the Lagrange function corresponds to a classical filtering problem. In this case, the matrices $N$, $T$ in (\ref{NS}), (\ref{TU}) vanish, and, under a simplifying assumption that the filter has the same state dimension $q=n$ as the plant, the necessary conditions of optimality (\ref{da0})--(\ref{de0}), (\ref{a}) take the form
\begin{align}
\label{da00}
    LM =& -I_n,\\
\label{a00}
    a
    =&
    -(LA+eC)M,\\
\label{db00}
    b =& 0,\\
\label{dc00}
    c =& (G^{\rT}G)^{-1} G^{\rT} F M,\\
\label{de00}
    e =& -L(P_{11}C^{\rT} + BD^{\rT}) - P_{21}C^{\rT}
     =
    -L(\Pi C^{\rT} + BD^{\rT}),\!
\end{align}
where, in view of (\ref{LM}), the real positive semi-definite symmetric matrix
\begin{equation}
\label{Pi}
    \Pi := P_{11} - MP_{21}
\end{equation}
is the Schur complement \cite{HJ_2007} of $P_{22}$ in (\ref{Pblocks}).
In particular, (\ref{db00}) shows that the additional noise $\omega$ (uncorrelated with the plant noise $w$) is redundant in the optimal filter, in conformance with the classical Kalman filtering theory.
Since (\ref{da00}) implies  that $L = -M^{-1}$, an appropriate similarity transformation $\xi\mapsto \sigma \xi$, $a\mapsto \sigma a\sigma^{-1}$, $b\mapsto \sigma b$, $c\mapsto c \sigma^{-1}$, $e\mapsto \sigma e$  of the filter (\ref{xi_eta}) with a nonsingular matrix $\sigma\in \mR^{n\x n}$ (which does not have to be symplectic in the sense that $\sigma \Theta_2 \sigma^{\rT} = \Theta_2$) leads to $L=-I_n$, $M = I_n$. The corresponding matrices $a$, $c$, $e$ in (\ref{a00}), (\ref{dc00}), (\ref{de00}) reduce to
\begin{equation}
\label{ace}
    a
    =
    A-eC,
    \quad
    c = (G^{\rT}G)^{-1} G^{\rT} F,
    \quad
    e = \Pi C^{\rT} + BD^{\rT},
\end{equation}
with $c$ being specified completely  by the matrices $F$, $G$. Accordingly,
the Lyapunov equation (\ref{PLyap11})--(\ref{PLyap22}) takes the form of an algebraic Riccati equation \cite{LR_1995} for the matrix $\Pi$ from (\ref{Pi}):
\begin{equation}
\label{RicPi}
    A\Pi + \Pi A^{\rT} + BB^{\rT} -
    (\Pi C^{\rT} + BD^{\rT}) (C\Pi  + DB^{\rT}) = 0.
\end{equation}
In view of (\ref{DDdd}), both (\ref{ace}) and (\ref{RicPi}) indeed correspond to the classical Kalman filter
SDE $
    \rd \xi = A\xi\rd t  + e(\rd y - C\xi \rd t)
$, with $\rd y - C\xi \rd t$  the innovation.

\appendix

\subsection{Proof of Lemma~\ref{lem:dEdabce}}\label{app:proof_lem:dEdabce}
%%%%%%%%%%%%%%%%%%%%%%%%%%%%%%%%%%%%%%%%%%%%%%%%%%%%%%%%%%%%%%%%%%%%%%%%%%%%%%%
\renewcommand{\theequation}{\Alph{subsection}\arabic{equation}}
\setcounter{equation}{0}

%==============================================================================
%\begin{proof}
Since the matrices $\cB$, $\cC$ do not depend on $a$, then, in view of (\ref{cAB}) and Lemma~\ref{lem:dEdABC}, the first variation of the function $\cE$ with respect to the matrix $a$ is
$
    \delta_a \cE
    =
    \BRA
        \d_{\cA} \cE,
        {\scriptsize\begin{bmatrix}
            0& 0\\
            0 & \delta a
        \end{bmatrix}}
    \KET
    =
    \bra H_{22}, \delta a\ket
$.
Hence,
%\begin{equation}
%\label{dEda}
$    \d_a \cE
    =
    H_{22}
$, %\end{equation}
which, in combination with $\d_a \bra \Xi \Theta_2, a\ket = \Xi \Theta_2$ from (\ref{cL}), implies (\ref{dLda}). Similarly, since $\cA$, $\cC$ do not depend on the matrix $b$, then, in view of (\ref{cAB}) and Lemma~\ref{lem:dEdABC}, the first variation of the function $\cE$ with respect to $b$ takes the form
$
    \delta_b \cE
    =
    \BRA
        \d_{\cB} \cE,
        {\scriptsize\begin{bmatrix}
            0& 0\\
            0 & \delta b
        \end{bmatrix}}
    \KET
    =
    \bra
        Q_{22}b,
        \delta b
    \ket
$,
so that
%\begin{equation}
%\label{dEdb}
$    \d_b \cE
    =
    Q_{22} b
$. %\end{equation}
In combination with $\d_b \bra \Xi, bJ_2 b^{\rT}\ket = -2\Xi bJ_2$ and $\d_b \bra \Gamma d J_2, b\ket = \Gamma d J_2$ from (\ref{cL}), this establishes (\ref{dLdb}). In a similar vein, since the matrices $\cA$, $\cB$ do not depend on $c$, then, in view of (\ref{cC}) and Lemma~\ref{lem:dEdABC}, the first variation of the function $\cE$ with respect to the matrix $c$ is computed as
$
    \delta_c \cE
    =
    \bra
        \d_{\cC} \cE,
        {\small\begin{bmatrix} 0 & -G\delta c\end{bmatrix}}
    \ket
    =
    -\bra
        G^{\rT}\cC P_{\bullet 2},
        \delta c
    \ket.
$
Therefore,
%\begin{equation}
%\label{dEdc}
$    \d_c \cE
    =
    -G^{\rT}\cC P_{\bullet 2}
    =
    -G^{\rT} F P_{12} + G^{\rT}G c P_{22}
$, %\end{equation}
which,   in combination with $\d_c \bra \Gamma^{\rT}\Theta_2, c\ket = \Gamma^{\rT}\Theta_2$ from (\ref{cL}), yields  (\ref{dLdc}). Finally, since $\cC$ in (\ref{cC}) is independent of the matrix $e$, then, in view of (\ref{cAB}) and Lemma~\ref{lem:dEdABC}, the first variation of the function $\cE$ with respect to $e$ is
$
    \delta_e \cE
    =
    \BRA
        \d_{\cA} \cE,
        {\scriptsize\begin{bmatrix}
            0 & 0\\
            \delta e C & 0
        \end{bmatrix}}
    \KET
    +
    \BRA
        \d_{\cB} \cE,
        {\scriptsize\begin{bmatrix}
            0 & 0\\
            \delta e D & 0
        \end{bmatrix}}
    \KET
     =
    \bra
        H_{21} C^{\rT}+(Q_{21} B + Q_{22} eD)D^{\rT},
        \delta e
    \ket
$,
and hence,
%\begin{equation}
%\label{dEde}
$    \d_e \cE
    =
    H_{21} C^{\rT}+Q_{21} BD^{\rT} + Q_{22} e%DD^{\rT}.
$, where use is made of (\ref{DDdd}). %\end{equation}
In view of $\d_e \bra \Xi, eDJ_1D^{\rT}e^{\rT}\ket = -2\Xi eDJ_1D^{\rT}$ in (\ref{cL}), this leads to (\ref{dLde}), thus completing the proof of the lemma. \hfill$\blacksquare$
%\end{proof}
%==================================================================================================

%%%%%%%%%%%%%%%%%%%%%%%%%%%%%%%%%%%%%%%%%%%%%%%%%%%%%%%%%%%%%%%%%%%%%%%%%%%%%%%
\subsection{Proof of Lemma~\ref{lem:QAP}}\label{app:proof_lem:QAP}
%%%%%%%%%%%%%%%%%%%%%%%%%%%%%%%%%%%%%%%%%%%%%%%%%%%%%%%%%%%%%%%%%%%%%%%%%%%%%%%
\renewcommand{\theequation}{\Alph{subsection}\arabic{equation}}
\setcounter{equation}{0}

%==================================================================================================
%\begin{proof}
Left multiplication of both sides of the Lyapunov equation (\ref{PLyap}) by $Q_{2\bullet}$ and right multiplication of the Lyapunov equation (\ref{QLyap}) by $P_{\bullet 2}$ yields
\begin{align}
\label{QAP10}
    Q_{2\bullet} \cA P + H_{2\bullet} \cA^{\rT} + Q_{2\bullet}\cB\cB^{\rT} & = 0,\\
\label{QAP20}
    \cA^{\rT} H_{\bullet 2} + Q \cA P_{\bullet 2} + \cC^{\rT}\cC P_{\bullet 2} & = 0.
\end{align}
Here, use is made of the identities $Q_{j\bullet} P = H_{j\bullet}$ and $Q P_{\bullet k}  = H_{\bullet k}$, which follow from the definition of the Hankelian $H$ in (\ref{dEdABC}) and its block partitioning as in (\ref{Pblocks}). The definition of the matrix $\cA$ in (\ref{cAB}) implies that
\begin{align}
\label{HA}
    H_{2\bullet} \cA^{\rT}
    & =
    {\small\begin{bmatrix}
        H_{21}A^{\rT} &
        H_{21} C^{\rT} e^{\rT} + H_{22} a^{\rT}
    \end{bmatrix}},\\
\label{AH}
    \cA^{\rT} H_{\bullet 2}
    & =
    {\small\begin{bmatrix}
        A^{\rT}H_{12} +  C^{\rT} e^{\rT} H_{22}\\
        a^{\rT}H_{22}
    \end{bmatrix}}.
\end{align}
By recalling the matrices $\cB$, $\cC$ from (\ref{cAB}), (\ref{cC}), and substituting
%\begin{align}
%\label{QBB}
$    (Q\cB\cB^{\rT})_{22}
     =
    (Q_{21} B + Q_{22} eD)D^{\rT}e^{\rT} + Q_{22}bb^{\rT}$ and %\\
%\label{CCP}
$    (\cC^{\rT}\cC P)_{22}
     =
    c^{\rT} G^{\rT} (Gc P_{22}-FP_{12})
$ %\end{align}
together with (\ref{DDdd}), (\ref{HA}), (\ref{AH}), into the appropriate blocks of (\ref{QAP10}), (\ref{QAP20}), the resulting equations lead to (\ref{QAP1}), (\ref{QAP2}). \hfill$\blacksquare$
%\end{proof}
%==================================================================================================

%%%%%%%%%%%%%%%%%%%%%%%%%%%%%%%%%%%%%%%%%%%%%%%%%%%%%%%%%%%%%%%%%%%%%%%%%%%%%%%
\subsection{Proof of Lemma~\ref{lem:sym}}\label{app:proof_lem:sym}
%%%%%%%%%%%%%%%%%%%%%%%%%%%%%%%%%%%%%%%%%%%%%%%%%%%%%%%%%%%%%%%%%%%%%%%%%%%%%%%
\renewcommand{\theequation}{\Alph{subsection}\arabic{equation}}
\setcounter{equation}{0}

%\begin{proof}
Substitution of the stationarity conditions (\ref{dLda0}), (\ref{dLdb0}), (\ref{dLde0}) and the PR conditions (\ref{filterPR1}), (\ref{filterPR2}) into the identity (\ref{QAP1}) yields
%\begin{align}
%\nonumber
%    &\Ups
%    +
%    \overbrace{\Xi eDJ_1D^{\rT}}^{\rm by\, (\ref{dLde0})} e^{\rT}
%     +
%     \overbrace{\Xi \Theta_2}^{\rm by\, (\ref{dLda0})} a^{\rT}
%     +
%     \overbrace{\Xi b J_2b^{\rT}
%     -
%     \Gamma c\Theta_2}^{\rm by\, (\ref{dLdb0}),\, (\ref{filterPR2})}\\
%\nonumber
%     & =
%     \Ups
%     +
%     \Xi
%     (\underbrace{eDJ_1D^{\rT} e^{\rT}
%        +
%        \Theta_2 a^{\rT} + b J_2b^{\rT}
%     }_{= -a\Theta_2\, {\rm by}\, (\ref{filterPR1})})- \Gamma c\Theta_2\\
%\label{QAP1new}
%     & =
%    \Ups  -
%    (\Xi a +\Gamma c)\Theta_2 = 0,
%\end{align}
\begin{align}
\nonumber
    &\Ups
    +
    \Xi eDJ_1D^{\rT} e^{\rT}
     +
     \Xi \Theta_2 a^{\rT}
     +
     \Xi b J_2b^{\rT}
     -
     \Gamma c\Theta_2\\
\nonumber
     & =
     \Ups
     +
     \Xi
     (eDJ_1D^{\rT} e^{\rT}
        +
        \Theta_2 a^{\rT} + b J_2b^{\rT}
     )- \Gamma c\Theta_2\\
\label{QAP1new}
     & =
    \Ups  -
    (\Xi a +\Gamma c)\Theta_2 = 0,
\end{align}
thus proving (\ref{QAP1eq}).
In a similar vein, by substituting (\ref{dLda0}), (\ref{dLdc0}) into (\ref{QAP2}), it follows that
\begin{equation}
\label{QAP2new}
    \Ups
    +
    a^{\rT} \Xi \Theta_2  - c^{\rT} \Gamma^{\rT}\Theta_2
    = \Ups
    +
    (a^{\rT} \Xi - c^{\rT} \Gamma^{\rT})\Theta_2 = 0.
\end{equation}
Since the left-hand sides of (\ref{QAP1new}), (\ref{QAP2new}) are equal to each other, and $\det\Theta_2\ne 0$, then  $\Xi a + \Gamma c = c^{\rT} \Gamma^{\rT} - a^{\rT} \Xi$. The latter equality, in view of   $\Xi^{\rT}=-\Xi$, is equivalent to the matrix $\Xi a + \Gamma c$ being symmetric. \hfill$\blacksquare$
%\end{proof}
%==================================================================================================

%%%%%%%%%%%%%%%%%%%%%%%%%%%%%%%%%%%%%%%%%%%%%%%%%%%%%%%%%%%%%%%%%%%%%%%%%%%%%%%
\subsection{Proof of Lemma~\ref{lem:rat}}\label{app:proof_lem:rat}
%%%%%%%%%%%%%%%%%%%%%%%%%%%%%%%%%%%%%%%%%%%%%%%%%%%%%%%%%%%%%%%%%%%%%%%%%%%%%%%
\renewcommand{\theequation}{\Alph{subsection}\arabic{equation}}
\setcounter{equation}{0}

%\begin{proof}
The equation (\ref{da0}) is obtained by the left multiplication of both sides of (\ref{dLda0}) by $Q_{22}^{-1}$ and right multiplication by $P_{22}^{-1}$, followed by using the identity $H_{22} = Q_{21} P_{12} + Q_{22}P_{22}$ together with the matrices $L$, $M$ from (\ref{LM}) and $N$, $S$ from (\ref{NS}). The equality (\ref{db0}) is obtained by the left multiplication of (\ref{dLdb0}) by $Q_{22}^{-1}$ and using the matrices $N$, $T$ from (\ref{NS}), (\ref{TU}). The equation (\ref{dc0}) is obtained by the left multiplication of (\ref{dLdc0})  by $(G^{\rT}G)^{-1}$ (secured by the full column rank of $G$)  and right multiplication by $P_{22}^{-1}$ and using the identity $\Gamma^{\rT}\Theta_2P_{22}^{-1} = (Q_{22}^{-1}\Gamma)^{\rT}Q_{22}\Theta_2P_{22}^{-1} = T^{\rT}U$ which follows from (\ref{TU}). The equation (\ref{de0}) is obtained by the left multiplication of (\ref{dLde0}) by $Q_{22}^{-1}$ and using the identity $H_{21} = Q_{21} P_{11} + Q_{22}P_{21}$ along with the matrices $L$, $N$ from (\ref{LM}), (\ref{NS}). Finally, (\ref{de0}) is obtained by the left multiplication of (\ref{QAP1eq}) by $Q_{22}^{-1}$ and right multiplication by $P_{22}^{-1}$ and using the representation $
    \Ups
    =
    Q_{22}(LAM + eC M + a)P_{22}
$ for the matrix $\Ups$ in (\ref{Q2AP2}) in terms of $L$, $M$ from (\ref{LM}). \hfill$\blacksquare$
\subsection{Positive  Semi-definiteness of Grade Two Operators}\label{app:two}
%%%%%%%%%%%%%%%%%%%%%%%%%%%%%%%%%%%%%%%%%%%%%%%%%%%%%%%%%%%%%%%%%%%%%%%%%%%%%%%
\renewcommand{\theequation}{\Alph{subsection}\arabic{equation}}
\setcounter{equation}{0}

%To this end, we will need a lemma on positive  definiteness of grade two operators. %%%%%%%%%%%%%%%%%%%%%%%%%%%%%%%%%%%%%%%%%%%%%%%%%%%%%%%%%%%%%%%%%%%%%%%%%%%%%%%%%%%%%%%%%%%%%%%%%%%
\begin{lem}%[on two operators]
\label{lem:two}
Suppose $\alpha\in \mS_r$, $\beta \in \mS_s$ and $\sigma\in \mA_r$, $\tau\in \mA_s$, with $\alpha\succ 0$, $\beta \succ 0$.   Then a criterion of positive semi-definiteness of the grade two special self-adjoint operator $\Phi:= \[[[\alpha, \beta \mid \sigma, \tau\]]]$, defined by (\ref{specoper}) on the Hilbert space $\mR^{r\x s}$, is given by
\begin{equation}
\label{Phipos}
    \Phi \succcurlyeq 0\
    \Longleftrightarrow\
    \br(\alpha^{-1}\sigma)
    \br(\tau \beta^{-1})\<1.
\end{equation}
\end{lem}\hfill$\square$
%%%%%%%%%%%%%%%%%%%%%%%%%%%%%%%%%%%%%%%%%%%%%%%%%%%%%%%%%%%%%%%%%%%%%%%%%%%%%%%%%%%%%%%%%%%%%%%%%%%%

\begin{proof}
Consider the decomposition of $\Phi$ into the sum $\Phi= \Psi +\mho$ of grade one self-adjoint operators $\Psi:= \[[[\alpha, \beta\]]]$ and $\mho := \[[[\sigma, \tau\]]]$, where $\Psi \succ 0$ due to the assumptions on the matrices $\alpha$, $\beta$. Hence,
$
    \Phi
    =
    \sqrt{\Psi}
    (I + \Delta)
    \sqrt{\Psi}
$,
where
$
    \Delta
    :=
    \Psi^{-1/2} \mho \Psi^{-1/2}
    =
    \[[[
        \alpha^{-1/2}\sigma\alpha^{-1/2},
        \beta^{-1/2}\tau\beta^{-1/2}
    \]]]
$
is a grade one self-adjoint operator whose spectrum is symmetric about the origin \cite[Lemma 1]{VP_2011a}. In view of this symmetry, the condition $\br(\Delta)\<1$ is not only sufficient, but is also necessary for the fulfillment of $\Phi\succcurlyeq 0$. It now remains to note \cite[Section 7]{VP_2011a} that
$
    \br(\Delta)
    =
    \br(\alpha^{-1/2}\sigma\alpha^{-1/2})
    \br(\beta^{-1/2}\tau\beta^{-1/2})
    =
    \br(\alpha^{-1}\sigma)
    \br(\tau\beta^{-1})
$,
and the equivalence (\ref{Phipos}) is proved. \end{proof}

\end{document}